\documentclass[trackchanges]{aastex7}

\usepackage{amsmath}
\usepackage{CJKutf8}
\usepackage{anyfontsize,graphicx}

\usepackage{CJKutf8}

\begin{document}
\begin{CJK*}{UTF8}{gbsn}

\title{First Detailed MeerKAT Imaging Spectroscopy of a Solar Flare}

\correspondingauthor{Yingjie Luo}
\email{Yingjie.Luo@glasgow.ac.uk}

\author[0000-0002-5431-545X]{Yingjie Luo \rm(骆英杰)}
\affiliation{School of Physics \& Astronomy,
University of Glasgow
G12 8QQ, Glasgow, UK}
\email{Yingjie.Luo@glasgow.ac.uk} 

\author[0000-0002-8078-0902]{Eduard P. Kontar}
\affiliation{School of Physics \& Astronomy,
University of Glasgow
G12 8QQ, Glasgow, UK}
\email{eduard.kontar@glasgow.ac.uk}

\author[0000-0002-0205-0808]{Roelf Du Toit Strauss}
\affiliation{Centre for Space Research, North-West University, Potchefstroom 2520, South Africa}
\affiliation{National Institute for Theoretical and Computational Sciences (NITheCS), North-West University, Potchefstroom 2520, South Africa}
\email{dutoit.strauss@gmail.com}

\author[0000-0002-5915-697X]{Gert J.~J.~Botha}
\affiliation{School of Engineering, Physics and Mathematics, Northumbria University, Newcastle upon Tyne, NE1 8ST, UK}
\email{gert.botha@northumbria.ac.uk}

\author[0000-0003-4142-366X]{Tomasz Mrozek}
\affiliation{Space Research Centre, Polish Academy of Sciences, ul. Bartycka 18a, 00-716 Warszawa, Poland}
\email{tmrozek@cbk.pan.wroc.pl}

\author[0000-0003-2846-2453]{Gelu M. Nita}
\affiliation{Center For Solar-Terrestrial Research, New Jersey Institute of Technology, Newark, NJ 07102, USA}
\email{gelu.m.nita@njit.edu}

\author[0000-0002-1691-0215]{Sarah Buchner}
\affiliation{South African Radio Astronomy Observatory, Liesbeek House, Cape Town, South Africa}
\email{sbuchner@sarao.ac.za}

\author[0000-0002-9875-7436]{James O. Chibueze}
\affiliation{UNISA Centre for Astrophysics and Space Sciences (UCASS), College of Science, Engineering and Technology, University of South Africa, Cnr Christian de Wet Rd and Pioneer Avenue, Florida 1709, P.O. Box 392, 0003 UNISA, South Africa}
\affiliation{Centre for Space Research, North-West University, Potchefstroom 2520, South Africa}
\email{chibujo@unisa.ac.za}

\accepted{2026 February 4}
 
\begin{abstract}

Radio observations provide powerful diagnostics of energy release, particle acceleration, and transport processes in solar flares. However, despite recent progress in radio interferometric imaging spectroscopy, current instruments still face limitations in image fidelity and resolution, restricting detailed spectroscopic studies of flaring regions. Here we present high-fidelity imaging spectroscopy of a M1.3 GOES class flare with MeerKAT, a precursor to the future-generation array SKA-Mid. 
Radio emissions at the observed frequencies typically originate in the low corona, offering valuable insights into magnetic reconnection and primary energy-release sites. The obtained images achieve an unprecedented dynamic range exceeding $10^3$, enabling simultaneous analysis of bright coherent bursts and faint incoherent emission from the active region. Multiple spatially distinct coherent sources are identified, implying contributions from different populations of accelerated electrons. The incoherent emission extends beyond AIA structures, highlighting MeerKAT's ability to detect dilute but hot plasma invisible to Extreme Ultraviolet  instruments. Combined with co-temporal Hard X-ray images and magnetic field extrapolations, the radio sources are located within distinct magnetic structures, further revealing their association with different populations of accelerated electrons. These results demonstrate MeerKAT imaging spectroscopy as powerful diagnostics of solar flares and pave the way for future solar flare studies with SKA-Mid.
\end{abstract}

\section{Introduction} \label{sec:intro}

Solar flares are the most explosive energy release events in the solar system, producing radiation across the entire electromagnetic spectrum from $\gamma$-rays to radio wavelengths 
\citep{2011SSRv..159....3D,2011SSRv..159...19F,2011SSRv..159..107H,2011SSRv..159..301K,2017LRSP...14....2B}. Radio observations, through a variety of emission mechanisms, provide crucial diagnostics of accelerated electrons, thermal plasma, and magnetic fields in the solar corona during flares 
\citep[][as reviews]{1985ARA&A..23..169D,1998ARA&A..36..131B,2011SSRv..159..225W,2023ARA&A..61..427G}. These diagnostics have been greatly advanced by the development of imaging spectroscopy with current radio interferometric arrays 
providing unprecedented time-frequency resolution imaging 
from decimetric
\citep[e.g.][]{2013ApJ...763L..21C,2021ApJ...911....4L,2022ApJ...940..137L,2024NatAs...8...50Y,2024ApJ...970...17S} to metric/decameteric wavelengths 
\citep[e.g.][]{2017NatCo...8.1515K,2020ApJ...893..115C,2021ApJ...917L..32C,2022ApJ...925..140G}. At decimetric wavelengths, radio emissions are particularly complex and variable: bright coherent bursts often dominate, 
while weaker incoherent components 
can also be present \citep{2017LRSP...14....2B}. 
Such emissions typically originate in the low corona—the most plausible site of primary energy release 
and particle acceleration during solar flares. 
For decimetric observations, achieving a sufficiently high dynamic range to simultaneously detect and analyze both intense coherent bursts and much weaker incoherent 
emission is critical \citep{2025SoPh..300...90B}. 
High dynamic range of up to 1500 has been achieved because these data were well suited to the self-calibration 
technique \citep[e.g.][]{1992ApJS...78..599W}
allowing to observe faint moving waves \citep{2005ApJ...620L..63W}.
Higher spatial and temporal resolution 
are likewise essential for resolving fine coronal 
structures and accurately tracing energy transport.

The Square Kilometre Array \citep[SKA;][]{2009IEEEP..97.1482D} is an upcoming next-generation interferometric array designed to meet the stringent requirements for high-fidelity radio imaging spectroscopy, which will help to address a number of solar physics challenges 
\citep{2019AdSpR..63.1404N}.  As the precursor to the mid-frequency component of SKA-Mid, MeerKAT is located in the Northern Cape province of South Africa \citep{2016mks..confE...1J}. The array comprises 64 antennas, each 13.5 m in diameter with a 3.8 m sub-reflector, providing excellent sensitivity. Forty-eight of these antennas are concentrated within a $\sim$1 km core, while the maximum baseline extends to $\sim$8 km. As part of the central array for SKA-Mid, MeerKAT's excellent \textit{uv} coverage ensures well-sampled visibilities and enables high-fidelity synthesis imaging. Although its baselines are shorter than the final 150~km configuration planned for SKA-Mid, MeerKAT nonetheless provides $8''(1\text{ GHz}/f)$ spatial resolution 
at the observing frequency $f$. MeerKAT currently operates in three frequency bands—UHF (544--1087 MHz), L (856--1711 MHz), and S (1750--3499 MHz). MeerKAT was originally designed for extragalactic and pulsar studies, and therefore requires the Sun to be placed outside the primary beam for safe solar observing in its current operational mode, though an engineering mode for on-boresight observations has recently been demonstrated \citep{2025FrASS..1266743K}. Solar flares nonetheless provide the most stringent test of MeerKAT's capabilities: flare emissions are orders of magnitude brighter than quiet Sun, evolve rapidly, and often involve multiple spatially distinct sources with diverse morphologies and emission mechanisms. Successfully imaging such events requires both high dynamic range and exceptional image fidelity.

In this work, we present the first detailed spectroscopic imaging of a solar flare with MeerKAT revealing spatially distinct coherent and incoherent radio sources with unprecedented fidelity. These observations represent a significant step forward for solar radio interferometry and demonstrate MeerKAT's unique capability to capture both intense coherent bursts and faint thermal emission within the same event.

This Letter is organized as follows. Section~\ref{sec: data} describes the GOES M1.3-class solar flare that occurred on 2024 December 29 and outlines the imaging procedure used to obtain high-fidelity images with a dynamic range exceeding 1000. Section~\ref{sec: anal} presents detailed analyses of the flare: Section~\ref{sec: vdy} examines the spatially resolved vector dynamic spectra of coherent sources, Section~\ref{sec: incoher} characterizes the incoherent emission, and Section~\ref{sec: magnetic} discusses the related coronal magnetic field configuration. Finally, Section~\ref{sec: conc} summarizes the results and their implications for future high-resolution solar observations with SKA-Mid.

\section{Observations and Radio Imaging Spectroscopy} \label{sec: data}

In this section, we present the GOES M1.3-class solar flare that occurred on 2024 December 29 around 11:20 UT, together with the MeerKAT imaging observations. As noted earlier, for this event, the Sun is positioned outside its main lobe for observations. During this observation, the pointing centers were set at angular distances of 2.25$^\circ$ (the closest safe separation for the L band), 2.40$^\circ$, 2.45$^\circ$, and 2.70$^\circ$ from the solar center, corresponding to four separate scans, each lasting 15 minutes. 
This Letter focuses on scan~4, during which multiple bright radio bursts were detected simultaneously with enhanced X-ray emission.

\subsection{Overview of the Event and Radio Observations} \label{sec:ovewview}

\begin{figure*}[!ht]
\centering
\includegraphics[width=\textwidth]{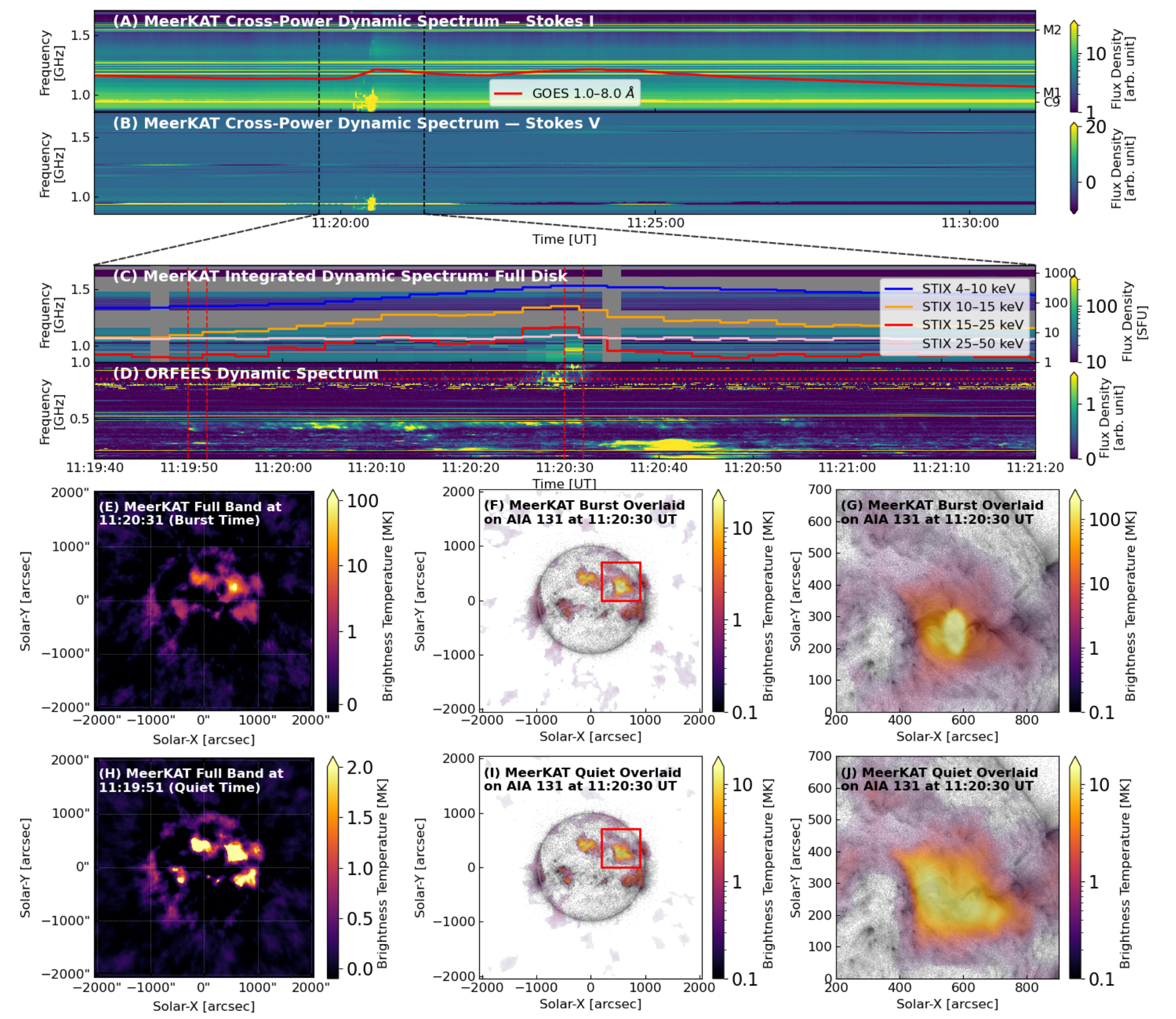}
\caption{Overview of the 2024 December 29 M1.3 GOES-class solar flare. 
(A) MeerKAT Stokes~I cross-power dynamic spectrum with GOES 1--8\,\AA\ soft X-ray flux (red) overplotted. 
(B) MeerKAT Stokes~V cross-power spectrum. 
(C) MeerKAT full-disk integrated dynamic spectrum during the burst with STIX light curves overlaid; bad channels and time intervals are flagged. 
(D) ORFEES dynamic spectrum. 
Vertical dashed black lines in (A) and (B) denote the interval shown in (C) and (D), while vertical dashed red lines in (C) and (D) mark the times used for imaging in (E--G) and (H--J). The red horizontal line in (D) indicates the low-frequency limit of the MeerKAT band. 
(E) MeerKAT full-band radio image in brightness temperature (MK) at the burst peak, a linear color scale for values below 1~MK and a logarithmic scale for values above 1~MK is used to enhance the visualization
(F) Alpha-blended overlay of MeerKAT emission (inferno colormap, transparency scaled by brightness) on an AIA 131\,\AA\ image (grayscale). Red boxes outline the regions shown in the zoomed-in view in (G). 
(H--J) Same as (E--G), but for the pre-burst quiet time.}

\label{fig: ova}
\end{figure*}

The GOES M1.3-class solar flare under study was located in AR~13936, a active region that earlier produced an GOES X1.1 flare at about 07:18~UT, accompanied by a coronal mass ejection (CME) on the same day. The M1.3 flare occurred in the northwestern portion of the solar disk, as indicated by the red box in Figure~\ref{fig: ova} (panels~F and~I). It took place during the decay phase of the prior flare, with the GOES 1--8\,\AA\ soft X-ray (SXR) flux peaking around 11:20~UT and showing a short duration (see the light curves in Figure~\ref{fig: ova}a). Hard X-ray (HXR) emission was also detected by the Spectrometer/Telescope for Imaging X-rays (STIX; \citealt{2020A&A...642A..15K}, see light curves in Figure~\ref{fig: ova}c) onboard Solar Orbiter (Sol-O; \citealt{2020A&A...642A...1M}), the Fermi Gamma-ray Burst Monitor (Fermi/GBM; \citealt{2009ApJ...702..791M}), and the Hard X-ray Imager (HXI; \citealt{2019RAA....19..163S}) onboard the Advanced Space-based Solar Observatory (ASO-S; \citealt{2019RAA....19..156G}).

MeerKAT captured this short-duration flare during a scan from 11:17:45.5 to 11:32:45.1~UT, with the phase center 
located 2.7$^\circ$ away from the solar disk. 
The observation utilized 62 of the 64 available antennas, achieving a maximum baseline length of 7698~m, 
corresponding to an angular resolution of about 8$''$ at 1~GHz. The observation was configured with 4096 frequency channels, providing sufficient spectral resolution for the analysis without generating excessive data volume, and achieved the highest temporal resolution available, 2~s. 
The data were recorded in full Stokes with linear polarization products (XX, XY, YX, and YY) in the L band, covering the frequency range 0.856--1.712~GHz. Figures~\ref{fig: ova}a and~\ref{fig: ova}b show the uncalibrated cross-power dynamic spectra in Stokes~I and~V, respectively, highlighting bright bursts and broadband emissions occurring near the peak of the GOES SXR and HXR bursts. The corrected flux spectrum from the solar disk, with details on imaging and primary-beam correction discussed in Section~\ref{sec: imaging}, is shown in Figure~\ref{fig: ova}c. For comparison, the dynamic spectrum from ORFEES \citep[Observations Radioélectriques pour FEDOME et l’Étude des Éruptions Solaires;][]{2021JSWSC..11...57H} is presented in panel~d, where overlapping frequency ranges exhibit 
similar features.

\subsection{Radio Imaging Spectroscopy} \label{sec: imaging}

We used the Common Astronomy Software Applications package (CASA; \citealt{2007ASPC..376..127M,2022PASP..134k4501C}) for data processing and imaging. The detailed data processing steps, from raw visibilities to the self-calibrated measurement set, are described in Appendix~\ref{app: a}. 
We employed the \texttt{tclean} task in CASA to perform synthesis imaging on the self-calibrated dataset. For the full-disk imaging, images were generated every 2~s (one temporal pixel) across 16 frequency channels ($\sim$ 3.34 MHz). For images containing bright coherent bursts, the dynamic range (DR)—defined as the ratio of the brightest pixel to the root-mean-square (rms) noise in a quiet region outside the solar disk—exceeded 1000. The maximum DR value, recorded around 0.96~GHz at 11:20:31~UT, reached 1482 (1037.1~MK / 0.7~MK). Images without bright bursts exhibited DRs between 60 and 300, with typical values above 120. Since the pointing center was located 2.7$^\circ$ off the solar disk, this DR does not reach the theoretical instrumental limit due to signal attenuation outside the main lobe. Nevertheless, it represents unprecedented performance 
at decimetric wavelengths for solar observations 
\citep{2025SoPh..300...90B}, for example, in studies 
using the VLA where dynamic ranges typically remained below 100 \citep[e.g.,][]{2021ApJ...911....4L}. Notably, the high fidelity of MeerKAT spectroscopic imaging is not only indicated by the high dynamic range, but more importantly by its ability to simultaneously recover clean, weak emission features in the presence of very bright sources.

To recover the correct solar flux density from the attenuated signal, we applied a primary-beam (PB) correction based on the array-averaged holography beam model \citep{2022AJ....163..135D,2023AJ....165...78D}. The correction was applied to every pixel of the image at each frequency channel. To avoid unrealistically large brightness temperatures in regions where the primary-beam response approaches a null and the signal level becomes comparable to or smaller than the noise level, an upper limit of 60 dB was imposed on the applied PB attenuation correction. This enables the construction of the full-disk integrated flux dynamic spectrum shown in Figure~\ref{fig: ova}c. Gray pixels denote flagged frequency channels and time intervals of bad data.

Further, to obtain a broadband composite image across the entire observed L band, intended to provide a representative broadband view of spatial morphology, we first converted each well-imaged (DR~$>$~50) sub-band Stokes~I image (in Jy~beam$^{-1}$) to specific intensity by dividing it by the synthesized-beam solid angle. We then obtained a frequency-averaged specific intensity ($I_\nu$) map over the usable frequency range and converted it to brightness temperature using the Rayleigh--Jeans relation. The composite image obtained at 11:20:31~UT during the flare peak is shown 
in Figure~\ref{fig: ova}e. The pixel associated with the brightest coherent burst reached a brightness temperature 
of 473~MK, while the composite image has a root-mean-square (rms) noise of about 0.11~MK in quiet regions, corresponding to a dynamic range exceeding 4000. 
The composite image was overlaid on the AIA 131~\AA\ image using a color--alpha blend, as shown in Figure~\ref{fig: ova}f--g. The resulting composite images in Figure~\ref{fig: ova}e--g clearly display, simultaneously, the bright bursts, the flare site, the major active regions on the disk, and the limb structures. In addition, these features align well with AIA 131~\AA\ emission. The only exception is the southwestern limb component, which appears less visible. 
This is primarily because the corresponding frequency band—expected to be least affected by primary-beam attenuation (near the sidelobe peak)—was flagged owing to radio-frequency interference (RFI), resulting in insufficient signal 
for detection.

We also applied the same imaging procedure to the pre-burst quiet period, as shown in Figure~\ref{fig: ova}h. The maximum brightness temperature reached $15.2$~MK, with an rms noise 
of $0.04$~MK. The synthetic image reveals a clearly identifiable flare site, and emission is confidently 
detected down to $0.1$~MK in brightness temperature (Figure~\ref{fig: ova}i--j). 
The extended features with relatively low brightness temperatures, visible in MeerKAT but absent in AIA, is likely outside the sensitivity range of AIA due to the lack sufficient emission measure for detection. This demonstrates that MeerKAT is capable of detecting thermal plasma invisible to EUV instruments. The images obtained at both narrow frequency sub-bands and across the broadband composite collectively show 
that MeerKAT provides unprecedented high-fidelity imaging that enables simultaneous analysis of bright coherent bursts, incoherent emission from the flare region, and faint extended emissions.

\section{Results and Analysis} \label{sec: anal}

In this section, we perform a detailed analysis of the different observed sources and investigate their 
associated physical processes.

\subsection{Spatially Distinct Coherent Radio Sources and Spatially Resolved Dynamic Spectra} \label{sec: vdy}

\begin{figure*}[!ht]
\centering
\includegraphics[width=\textwidth]{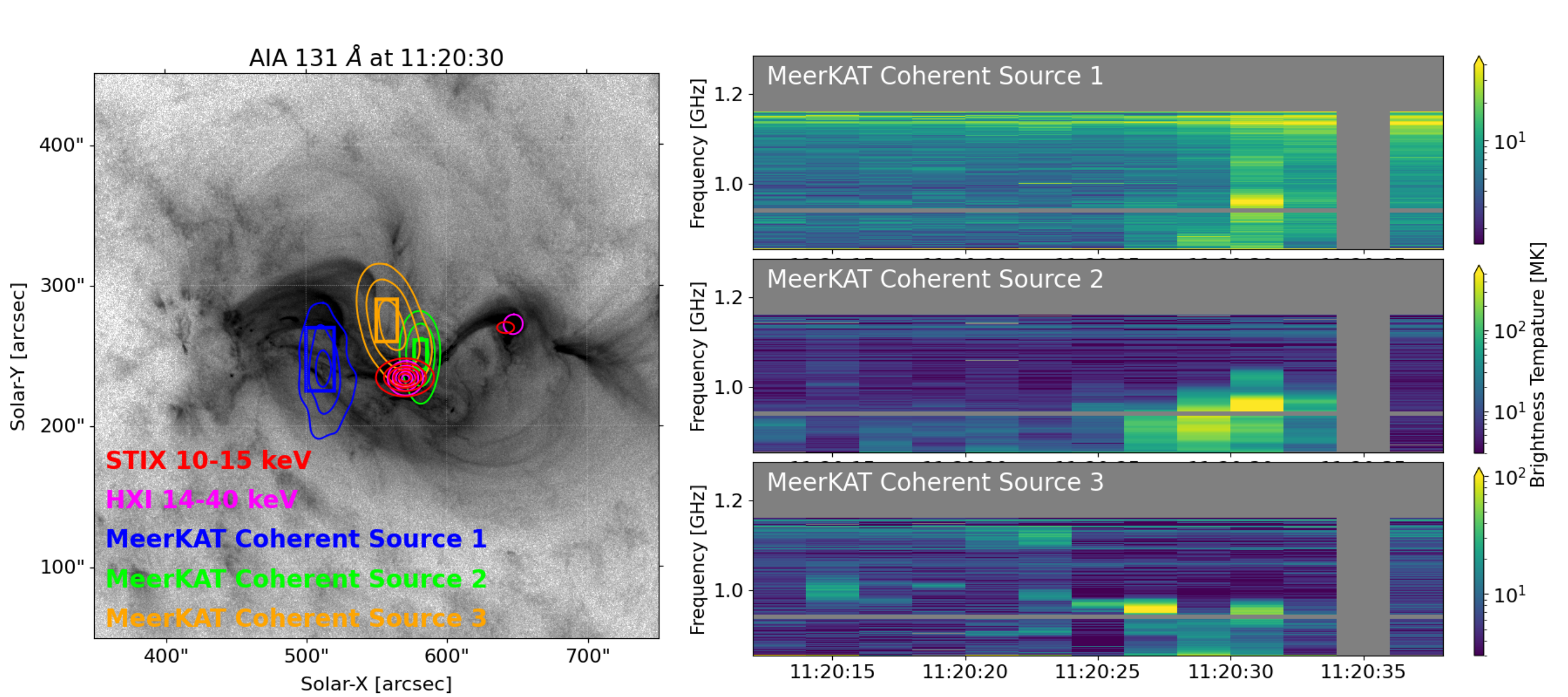}
\caption{Left: AIA 131\,\AA\ image of the flaring active region at 11:20:30\,UT, overlaid with different MeerKAT coherent sources (50, 70, and 90\% contours). HXI and STIX contours (10, 30, 50, 70, and 90\%) are also shown. Right: spatially resolved vector dynamic spectra of the selected sources (boxes in the left panel), obtained after subtracting contamination from nearby sources using two-dimensional Gaussian fitting. The images used to construct the vector dynamic spectrum are generated across four frequency channels, corresponding to a bandwidth of approximately 0.84 MHz.}
\label{fig: vdy}
\end{figure*}

During this short-duration flare, multiple radio emissions with varying properties were observed by MeerKAT. Based on their narrow bandwidths and high brightness temperatures, we identified three coherent sources located in spatially distinct regions (see Figure~\ref{fig: vdy}, left panel). Source~1 (blue contours) is located at a plausible loop leg 
of the southern loop system. The other two sources, Source~2 and Source~3 (green and orange contours, respectively), are situated close to each other near the center of the flare site. Source~2 lies closer to the HXR source, indicated by the magenta and red contours from HXI (14--40~keV) and STIX (10--15~keV), respectively, while Source~3 appears to be situated along another loop. The relationship between these sources and the magnetic field configuration will be discussed further in Section~\ref{sec: magnetic}.

We take advantage of MeerKAT's capability for high-fidelity spectroscopic imaging at each time-frequency pixel 
to generate four-dimensional image cubes (two spatial dimensions, one frequency dimension, and one time dimension) for each polarization product, focusing on Stokes~I in this study. These image cubes enable spatially resolved vector dynamic spectra, allowing detailed examination of the spectral-temporal properties of each source within the selected regions \citep{2015Sci...350.1238C}. 
In this event, Sources~2 and~3 are located in close proximity, resulting in partial flux leakage that affects their respective vector dynamic spectra. To accurately obtain the flux of the target source, we performed a two-dimensional Gaussian fit to the nearby source within the region enclosed by its 50\% contour and subtracted its contribution from the total flux. The resulting vector dynamic spectra for the different sources—calculated from the average brightness temperature within the boxed regions shown in the left panel of Figure~\ref{fig: vdy} (blue, green, and orange for Sources~1, 2, and~3, respectively)—are displayed in the right panels of Figure~\ref{fig: vdy}. The three sources exhibit distinct spectral characteristics. 
Source~1, although relatively weak, reaches a peak brightness temperature of 62~MK. Source~2, the brightest among the three, attains approximately 1100~MK and shows features that drift upward in frequency over a period of about 6~s (corresponding to three temporal pixels). 
Source~3 achieves a maximum brightness temperature 
of 170~MK, but its emission is only dominant 
within a very narrow frequency band around 11:20:26~UT.

\begin{figure*}[!ht]
\centering
\includegraphics[width=\textwidth]{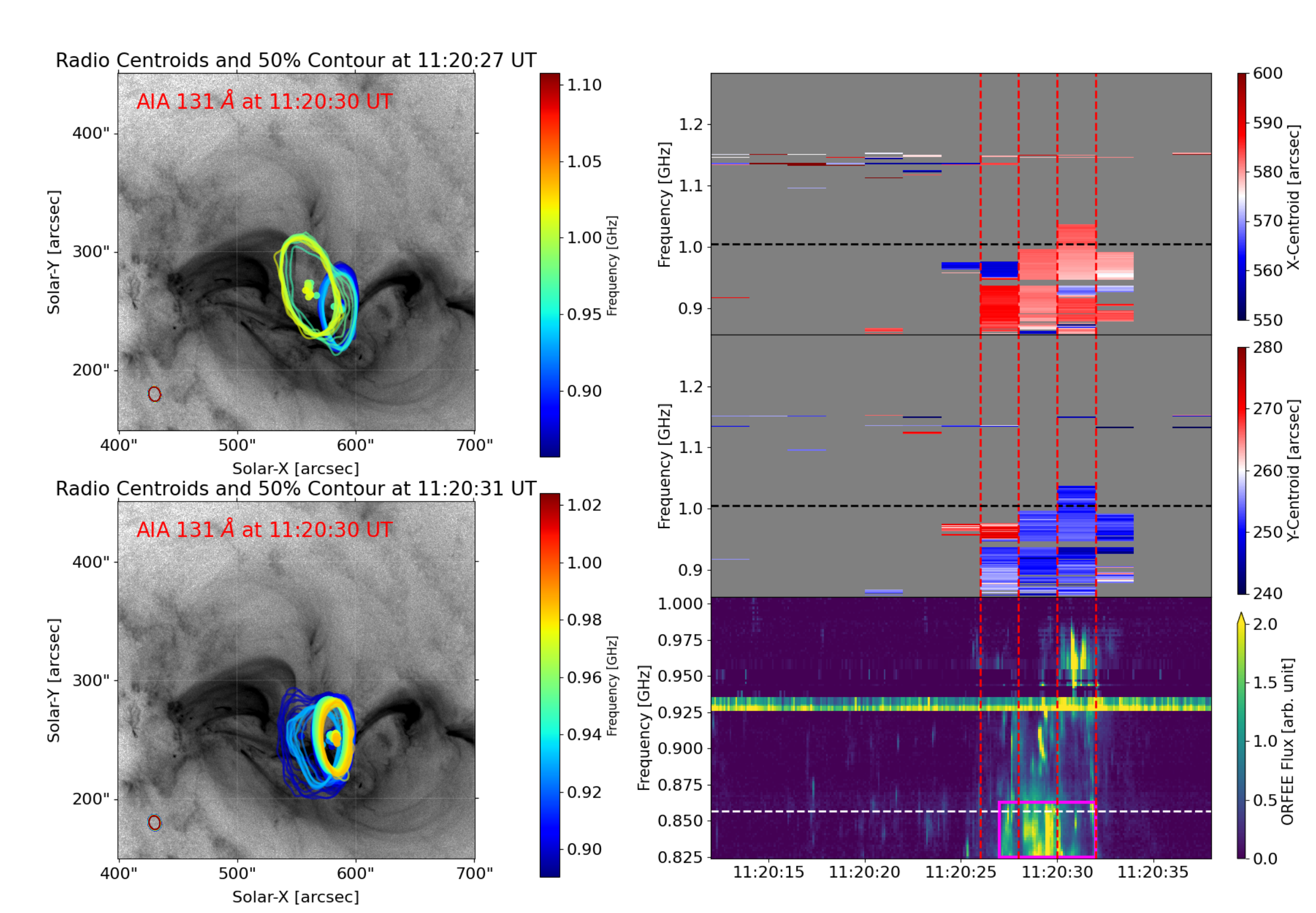}
\caption{Left panels: (Top) AIA 131\,\AA\ image of the flaring site at 11:20:30\,UT, overlaid with MeerKAT centroids (colored dots) and 50\% flux density contours (colored lines) at selected frequencies from 11:20:27\,UT. Colors denote the observing frequencies, and the synthesized beam used for the image is shown in the lower-left corner. Two spatially distinct sources are evident at different frequencies. (Bottom) Same as the top panel, but for 11:20:31\,UT. The selected frequencies differ, and source extensions appear at relatively lower frequencies. The extended low-frequency feature may be related to the structures indicated by the magenta box in the ORFEES dynamic spectrum, many of which fall below the MeerKAT L-band range. Uncertainties of the centroid positions were estimated as $\sigma \approx \theta_{\mathrm{beam}} / (\sqrt{8\ln2}\,\mathrm{SNR})$ \citep{1997PASP..109..166C}, and are smaller than the symbol size to be visualized. Right panels: (Top and middle) Temporal evolution of the X- and Y-centroid positions, showing only pixels with $T_{\mathrm{B}} > 30$\,MK. Black horizontal lines mark the upper limit of the ORFEES frequency coverage, and vertical red lines indicate the times corresponding to the left-panel plots. (Bottom) ORFEES dynamic spectrum during the same interval, with white horizontal lines marking the lower limit of the MeerKAT L-band coverage.
}
\label{fig: cens}
\end{figure*}

We further determined the source centroids by applying a second-order polynomial fit along orthogonal directions within a 10$\times$10 pixel area centered on the pixel with the maximum brightness temperature. The resulting centroid locations in Solar-X and Solar-Y coordinates are shown in the top and middle right panels of Figure~\ref{fig: cens}, respectively, where only pixels with maximum brightness temperatures above 30~MK are included. By examining the centroid spectra, we identify, at each time and frequency, which source dominates the emission (for example, Source~3 dominates around 0.96~GHz between 11:20:24 and 11:20:28~UT), thereby tracing the spatial and spectral evolution of the sources.

In the left panel of Figure~\ref{fig: cens}, we present the radio emissions at multiple frequencies, with their centroids and 50\% contours at two distinct time frames. 
At 11:20:27~UT, Sources~2 and~3 are clearly distinguishable, each dominating a different frequency range at that moment. Notably, these two sources do not coexist within any frequency band and are spatially separated without any smooth 
connecting features in either frequency or time, 
suggesting that they originate from distinct populations of accelerated electrons rather than from propagation 
of the same electron population. At 11:20:31~UT, the situation is different: at lower frequencies, Source~2 exhibits a smooth, frequency-dependent extension. 
This could be due to the transport of accelerated electrons; however, the 2~s temporal resolution limits our ability to investigate this further. 
To complement our analysis, we employed the ORFEES dynamic spectrum, which, although unable to image, provides a better temporal resolution of 0.1~s and offers additional insight (see Figure~\ref{fig: cens}, bottom-right panel). 
The extended low-frequency feature observed by MeerKAT is likely related to the structures seen in the ORFEES data, as highlighted by the magenta box in the bottom-right panel of Figure~\ref{fig: cens}. All these results suggest that accelerated electrons exist in multiple regions within the flare site, likely originating from different acceleration sites or resulting from energetic electron 
transport processes.

\subsection{Incoherent Emissions} \label{sec: incoher}

\begin{figure*}[!ht]
\centering
\includegraphics[width=\textwidth]{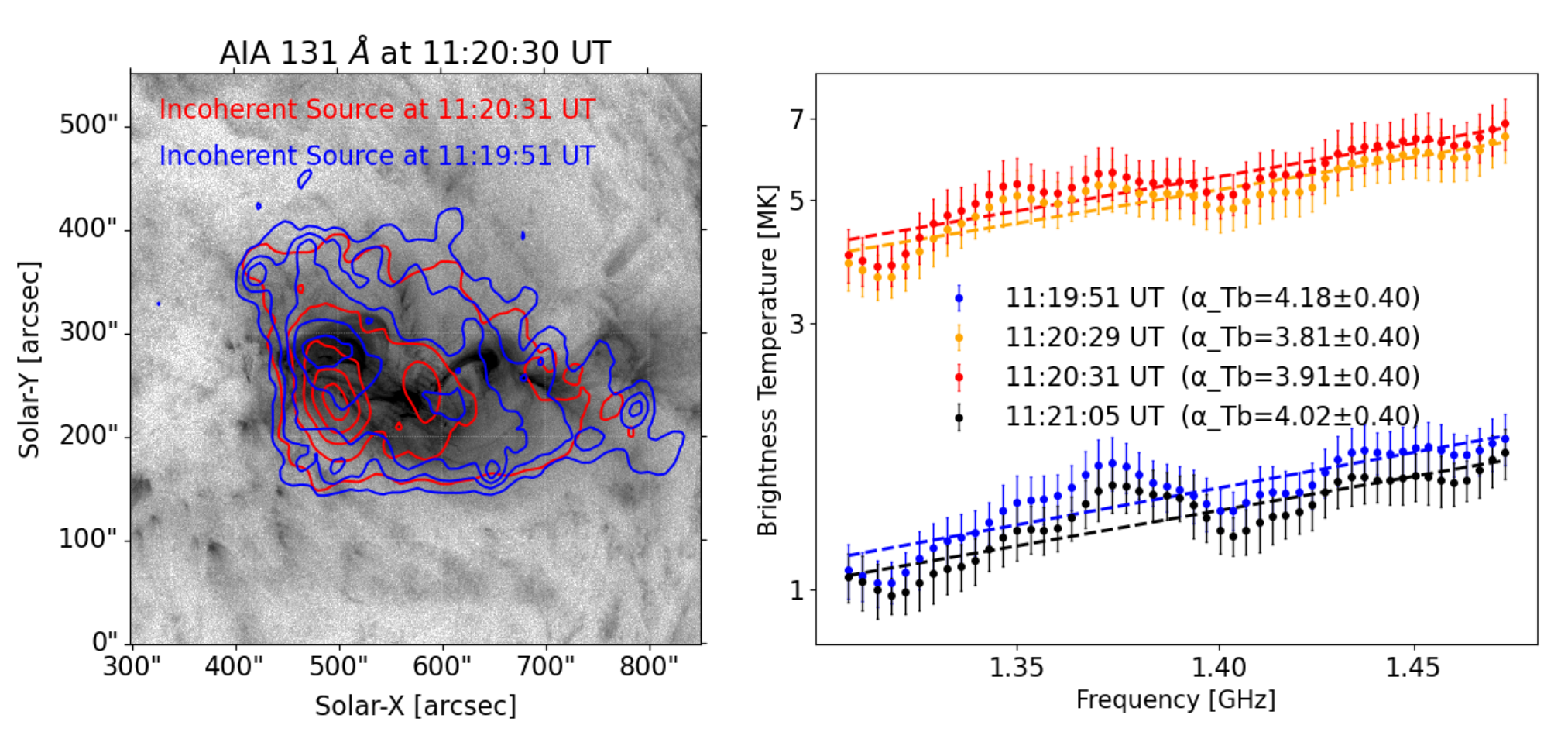}
\caption{Left: Contours (10, 30, 50, 70, and 90\%) of the incoherent radio source overlaid on the AIA 131\,\AA\ image at 11:20:30\,UT. Blue contours correspond to the pre-burst quiet time (11:19:51\,UT), and red contours to the burst time (11:20:31\,UT). Right: Brightness-temperature spectra of the incoherent source at four times: 11:19:51 (pre-burst), 11:20:29, 11:20:31, and 11:21:05\,UT (post-burst). Dashed lines show power-law fits obtained through weighted least-squares regression in logarithmic space, with the fitted spectral indices indicated in the legend. Uncertainties combine image rms (estimated from a quiet region) and PB correction errors (assumed to be 10\%). The spectra lie in the optically thick regime, exhibiting higher $T_{\mathrm{B}}$ and flatter indices during the burst.}
\label{fig: incor}
\end{figure*}

In addition to the coherent sources discussed in Section~\ref{sec: vdy}, we also examined the incoherent emission to investigate the properties of the flare plasma. As shown in Figure~\ref{fig: ova} (panels~G and~I) and further illustrated by the 10\% contours in the left panel of Figure~\ref{fig: incor}, the observed incoherent emission extends beyond the  loop structures visible in AIA images. 
This extended component is likely to be associated with faint free--free emission produced by less dense plasma at relatively higher altitudes above the visible flare loops. Although the plasma producing such emission has a small emission measure and would normally be difficult to detect, MeerKAT's high sensitivity and imaging dynamic range 
allow this faint component to be clearly observed before 
and during the flare.

We analyzed both the spectral and temporal characteristics of the incoherent emission. The $1.4$~GHz images during the coherent and HXR burst (red contours at 11:20:31~UT) and during the pre-burst quiet time (blue contours at 11:19:51~UT) are displayed in the left panel of Figure~\ref{fig: incor}. The 1.4~GHz emission at 11:20:31~UT clearly differs from the coherent emission observed at lower frequencies. It exhibits a spatial displacement of about 20$^{\prime\prime}$ in the maximum brightness-temperature pixel relative to the nearest coherent Source~1 centroid, a broadband and continuous spectral morphology, and a much lower peak brightness temperature of around $7$~MK, suggesting its incoherent nature.
The promptness and co-temporality with HXR emission suggests that this could be optically thick part of the gyrosynchrotron spectrum.

We also plotted the brightness-temperature spectra within the 90\% contours, representing the maximum values during the pre-burst (blue), burst (orange and red), and post-burst (black) phases, in the right panel of Figure~\ref{fig: incor}. The lower-frequency range is dominated by coherent emission, while several higher-frequency bands are affected by RFI, leaving only a narrow usable range for the incoherent emission spectrum. This limitation prevents a detailed spectral analysis, such as performing spectral fits to accurately determine the plasma properties. Nevertheless, the higher brightness temperatures and flatter power-law indices observed during the burst (derived from power-law fits in logarithmic space, corresponding to the optically thick regime) suggest rapid heating and subsequent cooling of the plasma on a timescale of less than 30~s. Studying the hot $>5$~MK plasma in the solar corona is often viewed as the key to understanding the heating in the solar active regions via impulsive nanoflare heating. However, non-equilibrium of the ionization \citep[][]{2021FrASS...8...33D} can make this hot plasma difficult to detect via line emissions because the ionization state of the plasma may lag behind the rapid temperature increases during a nanoflare.

These flare observation results demonstrate that MeerKAT is capable of detecting plasma emission that is insensitive to current EUV observations, and future observations with broader frequency coverage will allow more accurate and quantitative spectral diagnostics.

\subsection{Magnetic Field Configuration} \label{sec: magnetic}

\begin{figure*}[!ht]
\centering
\includegraphics[width=\textwidth]{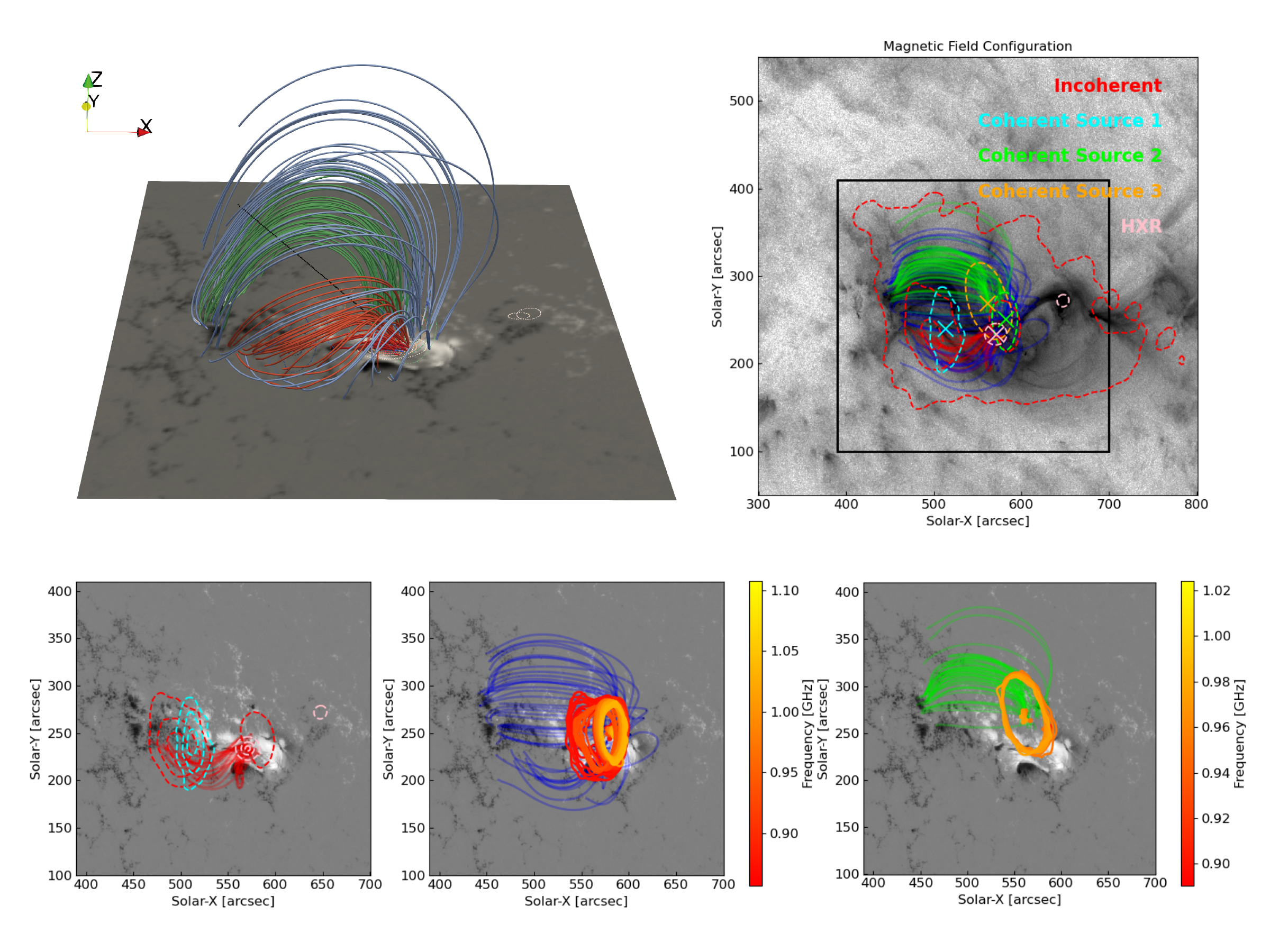}
\caption{Upper left: NLFFF magnetic field extrapolation showing a possible configuration of the observed sources. Black curves indicate the lines of sight from Earth. 
The $X$ and $Y$ axes correspond to heliographic longitude and latitude, while the $Z$ axis extends radially outward from the solar surface. 
Upper right: Magnetic field lines projected onto the helioprojective frame, with colored curves representing loops associated with different sources. Coherent radio sources are outlined by 50\% contours and marked by centroids in cyan, green, and orange, respectively. HXI sources are shown with 10\% contours and centroids in pink. Incoherent sources are outlined by 10\%, 50\%, and 90\% contours in red. 
Bottom panels: Magnetic field lines and associated sources within the field of view indicated by the black box in the upper-right panel. The left panel shows the HXR source (10\%, 50\%, and 90\% pink contours) together with coherent source~1 (50\%, 70\%, and 90\% cyan contours) and the incoherent sources (50\%, 70\%, and 90\% red contours) during the burst, while the middle and right panels display coherent Sources~2 and~3. Centroids in the middle and right panels are determined from different frequencies at 11:20:31~UT and 11:20:27~UT, respectively, corresponding to the times shown in the upper panel of Figure~\ref{fig: cens}. In the right panel, only the contours and centroids associated with Source~3 are shown.}
\label{fig: magn}
\end{figure*}

To further investigate the origin of the observed radio emissions, we reconstructed the coronal magnetic field to place the radio sources in their magnetic context. A nonlinear force-free magnetic field (NLFFF) extrapolation was performed using GX~Simulator \citep{2023ApJS..267....6N}, based on the SDO/HMI vector magnetogram taken at 11:10:30~UT. 
We identified magnetic loops likely associated with the observed sources, as illustrated in Figure~\ref{fig: magn}, where different colors indicate connections to various sources. The three-dimensional magnetic structure of these loops is shown in the upper-left panel, with the line of sight (LOS) from Earth’s perspective marked by a black line. 
The field lines projected onto the helioprojective coordinate system from this viewpoint are displayed in the upper-right panel of Figure~\ref{fig: magn}. It is important to note that the radio and HXR sources are observed along the LOS, meaning that they could, in principle, be located at any coronal height. Assuming the coherent sources are produced by plasma radiation, the ambient density profile can be used to constrain their possible heights. However, since these sources are located on the disk rather than near the limb—and given the dynamic plasma conditions during the flare—constraining the coronal density profile at the flare site remains challenging. Nevertheless, the observed frequencies (around 1~GHz, corresponding to an electron number density of $\sim1.2\times10^{10}$~cm$^{-3}$ derived from the plasma frequency relation $\nu_{\mathrm{pe}} \approx 8.98\times10^{3}\sqrt{n_{\mathrm{e}}}$~[Hz]) and their locations within 
the magnetic structure suggest that these sources 
are situated in the lower corona.

The bottom panels of Figure~\ref{fig: magn} illustrate representative magnetic field loops together with their likely associated sources in projection. In the bottom-left panel, the southern, lower-altitude loops associated with coherent Source~1 and the incoherent emission are shown. The HXR sources may also be linked to these loops but are located on the opposite side. The middle panel displays the larger, higher loops, where coherent Source~2—represented by multi-frequency centroids and the 50\% contour—appears to be situated at a loop leg. The multi-frequency 50\% contours extend along the loop toward lower frequencies. 
This behavior is consistent with plasma radiation, for which lower-frequency sources arise from regions of lower electron density, and therefore higher altitudes with larger loop cross-sections. These observations further suggest that Source~2 is associated with accelerated electrons distributed along the larger loop highlighted in blue. The right panel highlights the northern loops together with coherent Source~3. The distribution of multi-frequency centroids indicates that Source~3 is likely associated with trapped electrons within these loops. In addition, the loops containing Source~3 appear to lie beneath those connected to Source~2, consistent with the fact that Source~3 is observed at a higher frequency at 11:20:27~UT, corresponding to a higher electron density and a lower coronal height. Although the 2~s temporal resolution limits our ability to trace the detailed transport of accelerated electrons, the overall magnetic configuration confirms that the observed coherent sources originate from distinct regions associated with different populations of accelerated electrons.

\section{Discussion and Conclusions} \label{sec: conc}

In this Letter, we present the first detailed study of a GOES M1.3-class solar flare using high-fidelity MeerKAT imaging spectroscopy. With an unprecedented dynamic range exceeding 1000, our observations simultaneously capture bright coherent bursts, incoherent emissions from the flare site, and faint coronal features that remain undetected in AIA images. High-spatial- and spectral-resolution analyses of these emissions reveal several key results.

Three distinct coherent sources are identified, each exhibiting different spectral characteristics in the spatially resolved dynamic spectra. By locating their positions within the reconstructed magnetic field configuration, we find that they are likely associated with different coronal loops, indicating that the observed coherent bursts originate from accelerated electrons in distinct regions during the flare.

MeerKAT also detects faint emissions extending beyond the flare site that are invisible to SDO/AIA, highlighting its capability to diagnose plasma over a broader temperature range and with lower emission measures than those accessible to AIA. The spectra of these incoherent emissions during the burst show increased brightness temperatures and flatter spectral slopes, suggesting the presence of heated plasma followed by rapid cooling during this short-duration flare.

These findings demonstrate the potential of MeerKAT for solar flare studies using both coherent and incoherent radio emissions, and highlight the opportunities enabled by its high dynamic range and imaging spectroscopy capabilities. MeerKAT's combination of high spatial resolution and high dynamic range meets two essential requirements for diagnostics in the decimetric wavelength range—capabilities that remain limited in current radio arrays. Nevertheless, as a precursor rather than the final SKA-Mid system, 
MeerKAT still has inherent limitations that restrict further advances. The most significant limitation lies in the temporal resolution: most of the observed coherent bursts are not temporally resolved, making it difficult to trace the evolution of accelerated electrons responsible for the short-lived decimetric bursts. Notably, the coherent source exhibits an extended area exceeding 70$''$ at the 50\% contour, which could result from a superposition of multiple bursts within a 2~s integration, from the transport of accelerated electrons during this period, and/or from scattering effects. The limited temporal resolution thus prevents us from unambiguously determining which 
of these factors dominates. Future solar observations with MeerKAT, and ultimately with SKA-Mid, could benefit from improved temporal resolution, enabling more detailed studies of the rapid evolution of solar flare emission. The implementation of flexible observing modes, including higher-cadence sampling techniques applied in fast radio transient studies \citep{2025MNRAS.540.1685T}, could further enhance such capabilities. In addition, bore-sight observing modes would allow a more direct treatment of solar signal attenuation. Observations at other frequency bands, or the use of sub-array configurations to achieve broader instantaneous bandwidth, could also provide valuable complementary diagnostics. In addition, trigger-based or otherwise flexible observing strategies could further improve the efficiency of capturing such solar flare events. Once these challenges are addressed, MeerKAT and the future SKA-Mid will together usher in a new era of solar flare radio imaging-spectroscopy diagnostics.

\begin{acknowledgments}
The work was supported via the STFC/UKRI grants ST/T000422/1 and ST/Y001834/1. E.P.K. is supported by the Leverhulme Trust (Research Fellowship RF-2025-357). TM was funded  by the National Science Centre, Poland grant No. 2024/53/B/ST9/03028. GMN acknowledges support form NSF grants AGS-2425102 and RISE-2324724 to the New Jersey Institute of Technology. The MeerKAT telescope is operated by the South African Radio Astronomy Observatory, which is a facility of the National Research Foundation, an agency of the Department of Science, Technology and Innovation. The observations presented here formed part of the project ID: DDT-20241219-DS-01. Solar Orbiter is a space mission of international collaboration between ESA and NASA, operated by ESA. The STIX instrument is an international collaboration between Switzerland, Poland, France, Czech Republic, Germany, Austria, Ireland, and Italy. We acknowledge for the data resources from the Advanced Space-based Solar Observatory (ASO-S) mission and data service provided by National Space Science Data Center of China.

\end{acknowledgments}

\appendix

\section{Data processing: Partitioning, Flagging, Calibration, and Self-Calibration} \label{app: a}

This appendix describes the data processing from the raw visibilities to the self-calibrated measurement set ready for imaging. All processing was carried out using CASA. Data partitioning, flagging, and calibration procedures primarily followed the IDIA-\textsc{processMeerKAT} pipeline \citep{9560276}, with specific adjustments to key parameters to prevent solar emission—particularly during flaring intervals—from being inadvertently flagged or removed.

Data partitioning was performed in \texttt{mpiCASA} with multi-thread parallelization, producing smaller individual multi-measurement sets (MMSs) that facilitated faster processing. Initial flagging was carried out prior to calibration: we first manually flagged frequency channels affected by radio-frequency interference (RFI) and then applied the automated \texttt{tfcrop} algorithm using the \texttt{flagdata} task. The procedure was applied to all calibrator scans (J1939--6342 as the flux and bandpass calibrator, and J1830--3602 for amplitude and phase gain calibration, both standard celestial calibrators for MeerKAT) as well as to the solar scans. For the solar scans, the automatic flagging parameters were carefully tuned and subsequently verified to avoid mis-flagging the solar signal, especially when strong flaring activity was present.

After the initial flagging, standard flux calibration was applied to J1939--6342 using CASA task \texttt{setjy}. Subsequently, delay, bandpass, phase, and amplitude gain calibrations were performed on the appropriate calibrators in the first calibration round. After completing this round, additional flagging was conducted using both \texttt{tfcrop} and the more aggressive \texttt{rflag} algorithm, which is sensitive to bandpass shape. A second calibration round was then executed following the same procedure as the first. No further calibration rounds were required, as the resulting gain solutions were found to be close to unity, indicating no significant residual errors. Polarization calibration was not required for this dataset.

The calibrated measurement set was then phase-shifted to the solar disk center using the CASA task \texttt{phaseshift}, after which self-calibration was performed. The self-calibration was carried out individually for every 32 channels (corresponding to 6.7~MHz bandwidth). Three rounds of self-calibration were executed: two for phase and one for amplitude. The first round used a full-disk mask and generated the model using the CASA task \texttt{tclean}. The key \texttt{tclean} parameters were set to 5000 iterations (\texttt{niter=5000}) and Briggs weighting with a robustness parameter of 0.5. Baselines longer than 50~$\lambda$ were selected to ensure sufficient \textit{uv}-coverage sampling. The second round of phase self-calibration used a mask limited to the brightest active region or burst, which provided noticeably better results than reapplying the full-disk mask, especially for channels exhibiting bright bursts. In this round, the \texttt{tclean} parameters were adjusted to \texttt{niter=20000} and \texttt{robust=0}, including all available baselines. No further phase improvement was obtained beyond this stage, as additional rounds yielded no significant changes. The final amplitude self-calibration round used the same mask and parameters as the second round.

The self-calibration procedure led to observable, though moderate, improvements in imaging quality, with the dynamic range typically enhanced by a factor of 1.5--2.0 across different frequency bands. Such improvement is consistent with expectations, as the dataset was already well-calibrated before self-calibration. Over-self-calibration should be avoided, given the already excellent quality of the calibrated MeerKAT data. We note that the self-calibration process itself is inherently complex and requires careful evaluation at each stage. The parameters adopted here were thoroughly tested and optimized for this specific flare event, although they may need to be adjusted for other datasets.

\bibliography{sample701}{}

@ARTICLE{2020ApJ...893..115C,
       author = {{Chrysaphi}, Nicolina and {Reid}, Hamish A.~S. and {Kontar}, Eduard P.},
        title = "{First Observation of a Type II Solar Radio Burst Transitioning between a Stationary and Drifting State}",
      journal = {\apj},
         year = 2020,
        month = apr,
       volume = {893},
       number = {2},
          eid = {115},
        pages = {115},
          doi = {10.3847/1538-4357/ab80c1},
archivePrefix = {arXiv},
       eprint = {2003.11101},
 primaryClass = {astro-ph.SR},
       adsurl = {https://ui.adsabs.harvard.edu/abs/2020ApJ...893..115C}
}

@ARTICLE{2021ApJ...917L..32C,
       author = {{Clarkson}, Daniel L. and {Kontar}, Eduard P. and {Gordovskyy}, Mykola and {Chrysaphi}, Nicolina and {Vilmer}, Nicole},
        title = "{First Frequency-time-resolved Imaging Spectroscopy Observations of Solar Radio Spikes}",
      journal = {\apjl},
         year = 2021,
        month = aug,
       volume = {917},
       number = {2},
          eid = {L32},
        pages = {L32},
          doi = {10.3847/2041-8213/ac1a7d},
archivePrefix = {arXiv},
       eprint = {2108.06191},
 primaryClass = {astro-ph.SR},
       adsurl = {https://ui.adsabs.harvard.edu/abs/2021ApJ...917L..32C}
}

@ARTICLE{2022ApJ...925..140G,
       author = {{Gordovskyy}, Mykola and {Kontar}, Eduard P. and {Clarkson}, Daniel L. and {Chrysaphi}, Nicolina and {Browning}, Philippa K.},
        title = "{Sizes and Shapes of Sources in Solar Metric Radio Bursts}",
      journal = {\apj},
         year = 2022,
        month = feb,
       volume = {925},
       number = {2},
          eid = {140},
        pages = {140},
          doi = {10.3847/1538-4357/ac3bb7},
archivePrefix = {arXiv},
       eprint = {2111.07777},
 primaryClass = {astro-ph.SR},
       adsurl = {https://ui.adsabs.harvard.edu/abs/2022ApJ...925..140G}
}

@ARTICLE{1998ARA&A..36..131B,
   author = {{Bastian}, T.~S. and {Benz}, A.~O. and {Gary}, D.~E.},
    title = "{Radio Emission from Solar Flares}",
  journal = {\araa},
     year = 1998,
   volume = 36,
    pages = {131-188},
      doi = {10.1146/annurev.astro.36.1.131},
   adsurl = {http://adsabs.harvard.edu/abs/1998ARA%26A..36..131B}
}

@Article{2013ApJ...763L..21C,
  author        = {{Chen}, B. and {Bastian}, T.~S. and {White}, S.~M. and {Gary}, D.~E. and {Perley}, R. and {Rupen}, M. and {Carlson}, B.},
  journal       = {\apjl},
  title         = {{Tracing Electron Beams in the Sun's Corona with Radio Dynamic Imaging Spectroscopy}},
  year          = {2013},
  month         = jan,
  pages         = {L21},
  volume        = {763},
  adsurl        = {http://adsabs.harvard.edu/abs/2013ApJ...763L..21C},
  archiveprefix = {arXiv},
  doi           = {10.1088/2041-8205/763/1/L21},
  eid           = {L21},
  eprint        = {1211.3058},
  primaryclass  = {astro-ph.SR},
}

@ARTICLE{2017NatCo...8.1515K,
   author = {{Kontar}, E.~P. and {Yu}, S. and {Kuznetsov}, A.~A. and {Emslie}, A.~G. and
	{Alcock}, B. and {Jeffrey}, N.~L.~S. and {Melnik}, V.~N. and
	{Bian}, N.~H. and {Subramanian}, P.},
    title = "{Imaging spectroscopy of solar radio burst fine structures}",
  journal = {Nature Communications},
archivePrefix = "arXiv",
   eprint = {1708.06505},
 primaryClass = "astro-ph.SR",
     year = 2017,
    month = nov,
   volume = 8,
      eid = {1515},
    pages = {1515},
      doi = {10.1038/s41467-017-01307-8},
   adsurl = {http://adsabs.harvard.edu/abs/2017NatCo...8.1515K}
}

@ARTICLE{2019AdSpR..63.1404N,
   author = {{Nindos}, A. and {Kontar}, E.~P. and {Oberoi}, D.},
    title = "{Solar physics with the Square Kilometre Array}",
  journal = {Advances in Space Research},
archivePrefix = "arXiv",
   eprint = {1810.04951},
 primaryClass = "astro-ph.SR",
     year = 2019,
    month = feb,
   volume = 63,
    pages = {1404-1424},
      doi = {10.1016/j.asr.2018.10.023},
   adsurl = {https://ui.adsabs.harvard.edu/abs/2019AdSpR..63.1404N}
}

@ARTICLE{1992ApJS...78..599W,
       author = {{White}, S.~M. and {Kundu}, M.~R. and {Gopalswamy}, N.},
        title = "{High Dynamic Range Multifrequency Radio Observations of a Solar Active Region}",
      journal = {\apjs},
         year = 1992,
        month = feb,
       volume = {78},
        pages = {599},
          doi = {10.1086/191641},
       adsurl = {https://ui.adsabs.harvard.edu/abs/1992ApJS...78..599W}
}

@ARTICLE{2021FrASS...8...33D,
       author = {{Del Zanna}, Giulio and {Andretta}, Vincenzo and {Cargill}, Peter J. and {Corso}, Alain J. and {Daw}, Adrian N. and {Golub}, Leon and {Klimchuk}, James A. and {Mason}, Helen E.},
        title = "{High resolution soft X-ray spectroscopy and the quest for the hot (5-10 MK) plasma in solar active regions}",
      journal = {Frontiers in Astronomy and Space Sciences},
         year = 2021,
        month = apr,
       volume = {8},
          eid = {33},
        pages = {33},
          doi = {10.3389/fspas.2021.638489},
archivePrefix = {arXiv},
       eprint = {2103.06156},
 primaryClass = {astro-ph.SR},
       adsurl = {https://ui.adsabs.harvard.edu/abs/2021FrASS...8...33D}
}

@ARTICLE{2005ApJ...620L..63W,
       author = {{White}, S.~M. and {Thompson}, B.~J.},
        title = "{High-Cadence Radio Observations of an EIT Wave}",
      journal = {\apjl},
         year = 2005,
        month = feb,
       volume = {620},
       number = {1},
        pages = {L63-L66},
          doi = {10.1086/428428},
       adsurl = {https://ui.adsabs.harvard.edu/abs/2005ApJ...620L..63W}
}

@Article{2011SSRv..159....3D,
  Title                    = {Overview of the Volume},
  Author                   = {{Dennis}, B.~R. and {Emslie}, A.~G. and {Hudson}, H.~S.},
  Year                     = {2011},
  Doi                      = {10.1007/s11214-011-9802-z},
  Eprint                   = {1109.5831},
  Month                    = sep,
  Pages                    = {3-17},
  Volume                   = {159},
  Adsurl                   = {http://adsabs.harvard.edu/abs/2011SSRv..159....3D},
  Archiveprefix            = {arXiv},
  Journal                  = {\ssr},
  Primaryclass             = {astro-ph.SR}
}

@Article{2011SSRv..159..107H,
  Title                    = {Implications of X-ray Observations for Electron Acceleration and
 Propagation in Solar Flares},
  Author                   = {{Holman}, G.~D. and {Aschwanden}, M.~J. and {Aurass}, H. and {Battaglia},
 M. and {Grigis}, P.~C. and {Kontar}, E.~P. and {Liu}, W. and {Saint-Hilaire},
 P. and {Zharkova}, V.~V.},
  Year                     = {2011},
  Doi                      = {10.1007/s11214-010-9680-9},
  Eprint                   = {1109.6496},
  Month                    = sep,
  Pages                    = {107-166},
  Volume                   = {159},
  Adsurl                   = {http://adsabs.harvard.edu/abs/2011SSRv..159..107H},
  Archiveprefix            = {arXiv},
  Journal                  = {\ssr},
  Primaryclass             = {astro-ph.SR}
}

@Article{2011SSRv..159..301K,
  author        = {{Kontar}, E.~P. and {Brown}, J.~C. and {Emslie}, A.~G. and {Hajdas}, W. and {Holman}, G.~D. and {Hurford}, G.~J. and {Ka{\v s}parov{\'a}}, J. and {Mallik}, P.~C.~V. and {Massone}, A.~M. and {McConnell}, M.~L. and {Piana}, M. and {Prato}, M. and {Schmahl}, E.~J. and {Suarez-Garcia}, E.},
  journal       = {\ssr},
  title         = {Deducing Electron Properties from Hard X-ray Observations},
  year          = {2011},
  month         = sep,
  pages         = {301-355},
  volume        = {159},
  adsurl        = {http://adsabs.harvard.edu/abs/2011SSRv..159..301K},
  archiveprefix = {arXiv},
  doi           = {10.1007/s11214-011-9804-x},
  eprint        = {1110.1755},
  primaryclass  = {astro-ph.SR},
}

@Article{2011SSRv..159..225W,
  Title                    = {The Relationship Between Solar Radio and Hard X-ray Emission},
  Author                   = {{White}, S.~M. and {Benz}, A.~O. and {Christe}, S. and {F{\'a}rn{\'{\i}}k},
 F. and {Kundu}, M.~R. and {Mann}, G. and {Ning}, Z. and {Raulin},
 J.-P. and {Silva-V{\'a}lio}, A.~V.~R. and {Saint-Hilaire}, P. and
 {Vilmer}, N. and {Warmuth}, A.},
  Year                     = {2011},
  Doi                      = {10.1007/s11214-010-9708-1},
  Eprint                   = {1109.6629},
  Month                    = sep,
  Pages                    = {225-261},
  Volume                   = {159},
  Adsurl                   = {http://adsabs.harvard.edu/abs/2011SSRv..159..225W},
  Archiveprefix            = {arXiv},
  Journal                  = {\ssr},
  Primaryclass             = {astro-ph.SR}
}

@ARTICLE{2017LRSP...14....2B,
       author = {{Benz}, Arnold O.},
        title = "{Flare Observations}",
      journal = {Living Reviews in Solar Physics},
         year = 2017,
        month = dec,
       volume = {14},
       number = {1},
          eid = {2},
        pages = {2},
          doi = {10.1007/s41116-016-0004-3},
       adsurl = {https://ui.adsabs.harvard.edu/abs/2017LRSP...14....2B}
}

@ARTICLE{2011SSRv..159...19F,
       author = {{Fletcher}, L. and {Dennis}, B.~R. and {Hudson}, H.~S. and {Krucker}, S. and {Phillips}, K. and {Veronig}, A. and {Battaglia}, M. and {Bone}, L. and {Caspi}, A. and {Chen}, Q. and {Gallagher}, P. and {Grigis}, P.~T. and {Ji}, H. and {Liu}, W. and {Milligan}, R.~O. and {Temmer}, M.},
        title = "{An Observational Overview of Solar Flares}",
      journal = {\ssr},
         year = 2011,
        month = sep,
       volume = {159},
       number = {1-4},
        pages = {19-106},
          doi = {10.1007/s11214-010-9701-8},
archivePrefix = {arXiv},
       eprint = {1109.5932},
 primaryClass = {astro-ph.SR},
       adsurl = {https://ui.adsabs.harvard.edu/abs/2011SSRv..159...19F}
}

@ARTICLE{1985ARA&A..23..169D,
       author = {{Dulk}, G.~A.},
        title = "{Radio emission from the sun and stars.}",
      journal = {\araa},
         year = 1985,
        month = jan,
       volume = {23},
        pages = {169-224},
          doi = {10.1146/annurev.aa.23.090185.001125},
       adsurl = {https://ui.adsabs.harvard.edu/abs/1985ARA&A..23..169D}
}

@ARTICLE{2023ARA&A..61..427G,
       author = {{Gary}, Dale E.},
        title = "{New Insights from Imaging Spectroscopy of Solar Radio Emission}",
      journal = {\araa},
         year = 2023,
        month = aug,
       volume = {61},
        pages = {427-472},
          doi = {10.1146/annurev-astro-071221-052744},
       adsurl = {https://ui.adsabs.harvard.edu/abs/2023ARA&A..61..427G}
}

@ARTICLE{2021ApJ...911....4L,
       author = {{Luo}, Yingjie and {Chen}, Bin and {Yu}, Sijie and {Bastian}, T.~S. and {Krucker}, S{\"a}m},
        title = "{Radio Spectral Imaging of an M8.4 Eruptive Solar Flare: Possible Evidence of a Termination Shock}",
      journal = {\apj},
         year = 2021,
        month = apr,
       volume = {911},
       number = {1},
          eid = {4},
        pages = {4},
          doi = {10.3847/1538-4357/abe5a4},
archivePrefix = {arXiv},
       eprint = {2102.06259},
 primaryClass = {astro-ph.SR},
       adsurl = {https://ui.adsabs.harvard.edu/abs/2021ApJ...911....4L}
}

@ARTICLE{2015Sci...350.1238C,
       author = {{Chen}, Bin and {Bastian}, Timothy S. and {Shen}, Chengcai and {Gary}, Dale E. and {Krucker}, S{\"a}m and {Glesener}, Lindsay},
        title = "{Particle acceleration by a solar flare termination shock}",
      journal = {Science},
         year = 2015,
        month = dec,
       volume = {350},
       number = {6265},
        pages = {1238-1242},
          doi = {10.1126/science.aac8467},
archivePrefix = {arXiv},
       eprint = {1512.02237},
 primaryClass = {astro-ph.SR},
       adsurl = {https://ui.adsabs.harvard.edu/abs/2015Sci...350.1238C}
}

@ARTICLE{2022ApJ...940..137L,
       author = {{Luo}, Yingjie and {Chen}, Bin and {Yu}, Sijie and {Battaglia}, Marina and {Sharma}, Rohit},
        title = "{Multiple Regions of Nonthermal Quasiperiodic Pulsations during the Impulsive Phase of a Solar Flare}",
      journal = {\apj},
         year = 2022,
        month = dec,
       volume = {940},
       number = {2},
          eid = {137},
        pages = {137},
          doi = {10.3847/1538-4357/ac997a},
archivePrefix = {arXiv},
       eprint = {2210.06219},
 primaryClass = {astro-ph.SR},
       adsurl = {https://ui.adsabs.harvard.edu/abs/2022ApJ...940..137L}
}

@ARTICLE{2024ApJ...970...17S,
       author = {{Sharma}, Rohit and {Battaglia}, Marina and {Yu}, Sijie and {Chen}, Bin and {Luo}, Yingjie and {Krucker}, S{\"a}m},
        title = "{Study of Particle Acceleration Using Fine Structures and Oscillations in Microwaves from the Electron Cyclotron Maser}",
      journal = {\apj},
         year = 2024,
        month = jul,
       volume = {970},
       number = {1},
          eid = {17},
        pages = {17},
          doi = {10.3847/1538-4357/ad4884},
archivePrefix = {arXiv},
       eprint = {2405.04351},
 primaryClass = {astro-ph.SR},
       adsurl = {https://ui.adsabs.harvard.edu/abs/2024ApJ...970...17S}
}

@ARTICLE{2024NatAs...8...50Y,
       author = {{Yu}, Sijie and {Chen}, Bin and {Sharma}, Rohit and {Bastian}, Timothy S. and {Mondal}, Surajit and {Gary}, Dale E. and {Luo}, Yingjie and {Battaglia}, Marina},
        title = "{Detection of long-lasting aurora-like radio emission above a sunspot}",
      journal = {Nature Astronomy},
         year = 2024,
        month = jan,
       volume = {8},
       number = {1},
        pages = {50-59},
          doi = {10.1038/s41550-023-02122-6},
archivePrefix = {arXiv},
       eprint = {2310.01240},
 primaryClass = {astro-ph.SR},
       adsurl = {https://ui.adsabs.harvard.edu/abs/2024NatAs...8...50Y}
}

@ARTICLE{2025SoPh..300...90B,
       author = {{Bastian}, Timothy and {Chen}, Bin and {Mondal}, Surajit and {Saint-Hilaire}, Pascal},
        title = "{Noise in Maps of the Sun at Radio Wavelengths II: Solar Use Cases}",
      journal = {\solphys},
         year = 2025,
        month = jul,
       volume = {300},
       number = {7},
          eid = {90},
        pages = {90},
          doi = {10.1007/s11207-025-02498-w},
archivePrefix = {arXiv},
       eprint = {2506.09843},
 primaryClass = {astro-ph.SR},
       adsurl = {https://ui.adsabs.harvard.edu/abs/2025SoPh..300...90B}
}

@INPROCEEDINGS{2016mks..confE...1J,
       author = {{Jonas}, J. and {MeerKAT Team}},
        title = "{The MeerKAT Radio Telescope}",
    booktitle = {MeerKAT Science: On the Pathway to the SKA},
         year = 2016,
        month = jan,
          eid = {1},
        pages = {1},
          doi = {10.22323/1.277.0001},
       adsurl = {https://ui.adsabs.harvard.edu/abs/2016mks..confE...1J}
}

@ARTICLE{2009IEEEP..97.1482D,
       author = {{Dewdney}, P.~E. and {Hall}, P.~J. and {Schilizzi}, R.~T. and {Lazio}, T.~J.~L.~W.},
        title = "{The Square Kilometre Array}",
      journal = {IEEE Proceedings},
         year = 2009,
        month = aug,
       volume = {97},
       number = {8},
        pages = {1482-1496},
          doi = {10.1109/JPROC.2009.2021005},
       adsurl = {https://ui.adsabs.harvard.edu/abs/2009IEEEP..97.1482D}
}

@ARTICLE{2020A&A...642A..15K,
       author = {{Krucker}, S{\"a}m and {Hurford}, G.~J. and {Grimm}, O. and {K{\"o}gl}, S. and {Gr{\"o}belbauer}, H. -P. and {Etesi}, L. and {Casadei}, D. and {Csillaghy}, A. and {Benz}, A.~O. and {Arnold}, N.~G. and {Molendini}, F. and {Orleanski}, P. and {Schori}, D. and {Xiao}, H. and {Kuhar}, M. and {Hochmuth}, N. and {Felix}, S. and {Schramka}, F. and {Marcin}, S. and {Kobler}, S. and {Iseli}, L. and {Dreier}, M. and {Wiehl}, H.~J. and {Kleint}, L. and {Battaglia}, M. and {Lastufka}, E. and {Sathiapal}, H. and {Lapadula}, K. and {Bednarzik}, M. and {Birrer}, G. and {Stutz}, St. and {Wild}, Ch. and {Marone}, F. and {Skup}, K.~R. and {Cichocki}, A. and {Ber}, K. and {Rutkowski}, K. and {Bujwan}, W. and {Juchnikowski}, G. and {Winkler}, M. and {Darmetko}, M. and {Michalska}, M. and {Seweryn}, K. and {Bia{\l}ek}, A. and {Osica}, P. and {Sylwester}, J. and {Kowalinski}, M. and {{\'S}cis{\l}owski}, D. and {Siarkowski}, M. and {St{\k{e}}{\'s}licki}, M. and {Mrozek}, T. and {Podg{\'o}rski}, P. and {Meuris}, A. and {Limousin}, O. and {Gevin}, O. and {Le Mer}, I. and {Brun}, S. and {Strugarek}, A. and {Vilmer}, N. and {Musset}, S. and {Maksimovi{\'c}}, M. and {F{\'a}rn{\'\i}k}, F. and {Koz{\'a}{\v{c}}ek}, Z. and {Ka{\v{s}}parov{\'a}}, J. and {Mann}, G. and {{\"O}nel}, H. and {Warmuth}, A. and {Rendtel}, J. and {Anderson}, J. and {Bauer}, S. and {Dionies}, F. and {Paschke}, J. and {Pl{\"u}schke}, D. and {Woche}, M. and {Schuller}, F. and {Veronig}, A.~M. and {Dickson}, E.~C.~M. and {Gallagher}, P.~T. and {Maloney}, S.~A. and {Bloomfield}, D.~S. and {Piana}, M. and {Massone}, A.~M. and {Benvenuto}, F. and {Massa}, P. and {Schwartz}, R.~A. and {Dennis}, B.~R. and {van Beek}, H.~F. and {Rodr{\'\i}guez-Pacheco}, J. and {Lin}, R.~P.},
        title = "{The Spectrometer/Telescope for Imaging X-rays (STIX)}",
      journal = {\aap},
         year = 2020,
        month = oct,
       volume = {642},
          eid = {A15},
        pages = {A15},
          doi = {10.1051/0004-6361/201937362},
       adsurl = {https://ui.adsabs.harvard.edu/abs/2020A&A...642A..15K}
}

@ARTICLE{2019RAA....19..163S,
       author = {{Su}, Yang and {Liu}, Wei and {Li}, You-Ping and {Zhang}, Zhe and {Hurford}, Gordon J. and {Chen}, Wei and {Huang}, Yu and {Li}, Zhen-Tong and {Jiang}, Xian-Kai and {Wang}, Hao-Xiang and {Xia}, Fan-Xiao-Yu and {Chen}, Chang-Xue and {Yu}, Wen-Hui and {Yu}, Fu and {Wu}, Jian and {Gan}, Wei-Qun},
        title = "{Simulations and software development for the Hard X-ray Imager onboard ASO-S}",
      journal = {Research in Astronomy and Astrophysics},
         year = 2019,
        month = nov,
       volume = {19},
       number = {11},
          eid = {163},
        pages = {163},
          doi = {10.1088/1674-4527/19/11/163},
       adsurl = {https://ui.adsabs.harvard.edu/abs/2019RAA....19..163S}
}

@ARTICLE{2019RAA....19..156G,
       author = {{Gan}, Wei-Qun and {Zhu}, Cheng and {Deng}, Yuan-Yong and {Li}, Hui and {Su}, Yang and {Zhang}, Hai-Ying and {Chen}, Bo and {Zhang}, Zhe and {Wu}, Jian and {Deng}, Lei and {Huang}, Yu and {Yang}, Jian-Feng and {Cui}, Ji-Jun and {Chang}, Jin and {Wang}, Chi and {Wu}, Ji and {Yin}, Zeng-Shan and {Chen}, Wen and {Fang}, Cheng and {Yan}, Yi-Hua and {Lin}, Jun and {Xiong}, Wei-Ming and {Chen}, Bin and {Bao}, Hai-Chao and {Cao}, Cai-Xia and {Bai}, Yan-Ping and {Wang}, Tao and {Chen}, Bing-Long and {Li}, Xin-Yu and {Zhang}, Ye and {Feng}, Li and {Su}, Jiang-Tao and {Li}, Ying and {Chen}, Wei and {Li}, You-Ping and {Su}, Ying-Na and {Wu}, Hai-Yan and {Gu}, Mei and {Huang}, Lei and {Tang}, Xue-Jun},
        title = "{Advanced Space-based Solar Observatory (ASO-S): an overview}",
      journal = {Research in Astronomy and Astrophysics},
         year = 2019,
        month = nov,
       volume = {19},
       number = {11},
          eid = {156},
        pages = {156},
          doi = {10.1088/1674-4527/19/11/156},
       adsurl = {https://ui.adsabs.harvard.edu/abs/2019RAA....19..156G}
}

@ARTICLE{2020A&A...642A...1M,
       author = {{M{\"u}ller}, D. and {St. Cyr}, O.~C. and {Zouganelis}, I. and {Gilbert}, H.~R. and {Marsden}, R. and {Nieves-Chinchilla}, T. and {Antonucci}, E. and {Auch{\`e}re}, F. and {Berghmans}, D. and {Horbury}, T.~S. and {Howard}, R.~A. and {Krucker}, S. and {Maksimovic}, M. and {Owen}, C.~J. and {Rochus}, P. and {Rodriguez-Pacheco}, J. and {Romoli}, M. and {Solanki}, S.~K. and {Bruno}, R. and {Carlsson}, M. and {Fludra}, A. and {Harra}, L. and {Hassler}, D.~M. and {Livi}, S. and {Louarn}, P. and {Peter}, H. and {Sch{\"u}hle}, U. and {Teriaca}, L. and {del Toro Iniesta}, J.~C. and {Wimmer-Schweingruber}, R.~F. and {Marsch}, E. and {Velli}, M. and {De Groof}, A. and {Walsh}, A. and {Williams}, D.},
        title = "{The Solar Orbiter mission. Science overview}",
      journal = {\aap},
         year = 2020,
        month = oct,
       volume = {642},
          eid = {A1},
        pages = {A1},
          doi = {10.1051/0004-6361/202038467},
archivePrefix = {arXiv},
       eprint = {2009.00861},
 primaryClass = {astro-ph.SR},
       adsurl = {https://ui.adsabs.harvard.edu/abs/2020A&A...642A...1M}
}

@ARTICLE{2009ApJ...702..791M,
       author = {{Meegan}, Charles and {Lichti}, Giselher and {Bhat}, P.~N. and {Bissaldi}, Elisabetta and {Briggs}, Michael S. and {Connaughton}, Valerie and {Diehl}, Roland and {Fishman}, Gerald and {Greiner}, Jochen and {Hoover}, Andrew S. and {van der Horst}, Alexander J. and {von Kienlin}, Andreas and {Kippen}, R. Marc and {Kouveliotou}, Chryssa and {McBreen}, Sheila and {Paciesas}, W.~S. and {Preece}, Robert and {Steinle}, Helmut and {Wallace}, Mark S. and {Wilson}, Robert B. and {Wilson-Hodge}, Colleen},
        title = "{The Fermi Gamma-ray Burst Monitor}",
      journal = {\apj},
         year = 2009,
        month = sep,
       volume = {702},
       number = {1},
        pages = {791-804},
          doi = {10.1088/0004-637X/702/1/791},
archivePrefix = {arXiv},
       eprint = {0908.0450},
 primaryClass = {astro-ph.IM},
       adsurl = {https://ui.adsabs.harvard.edu/abs/2009ApJ...702..791M}
}

@ARTICLE{2021JSWSC..11...57H,
       author = {{Hamini}, Abdallah and {Auxepaules}, Gabriel and {Bir{\'e}e}, Lionel and {Kenfack}, Guy and {Kerdraon}, Alain and {Klein}, Karl-Ludwig and {Lespagnol}, Patrice and {Masson}, Sophie and {Coutouly}, Lucile and {Fabrice}, Christian and {Romagnan}, Renaud},
        title = "{ORFEES - a radio spectrograph for the study of solar radio bursts and space weather applications}",
      journal = {Journal of Space Weather and Space Climate},
         year = 2021,
        month = oct,
       volume = {11},
          eid = {57},
        pages = {57},
          doi = {10.1051/swsc/2021039},
       adsurl = {https://ui.adsabs.harvard.edu/abs/2021JSWSC..11...57H}
}

@INPROCEEDINGS{9560276,
  author={Collier, Jordan D. and Frank, Bradley and Sekhar, Srikrishna and Taylor, Andrew R.},
  booktitle={2021 XXXIVth General Assembly and Scientific Symposium of the International Union of Radio Science (URSI GASS)}, 
  title={The IDIA PROCESSMEERKAT pipeline: Fast CASA Processing on a Cloud-Based HPC Cluster}, 
  year={2021},
  volume={},
  number={},
  pages={1-4},
  doi={10.23919/URSIGASS51995.2021.9560276}}

@INPROCEEDINGS{2007ASPC..376..127M,
       author = {{McMullin}, J.~P. and {Waters}, B. and {Schiebel}, D. and {Young}, W. and {Golap}, K.},
        title = "{CASA Architecture and Applications}",
    booktitle = {Astronomical Data Analysis Software and Systems XVI},
         year = 2007,
       editor = {{Shaw}, R.~A. and {Hill}, F. and {Bell}, D.~J.},
       series = {Astronomical Society of the Pacific Conference Series},
       volume = {376},
        month = oct,
        pages = {127},
       adsurl = {https://ui.adsabs.harvard.edu/abs/2007ASPC..376..127M}
}

@ARTICLE{2022PASP..134k4501C,
       author = {{CASA Team} and {Bean}, Ben and {Bhatnagar}, Sanjay and {Castro}, Sandra and {Donovan Meyer}, Jennifer and {Emonts}, Bjorn and {Garcia}, Enrique and {Garwood}, Robert and {Golap}, Kumar and {Gonzalez Villalba}, Justo and {Harris}, Pamela and {Hayashi}, Yohei and {Hoskins}, Josh and {Hsieh}, Mingyu and {Jagannathan}, Preshanth and {Kawasaki}, Wataru and {Keimpema}, Aard and {Kettenis}, Mark and {Lopez}, Jorge and {Marvil}, Joshua and {Masters}, Joseph and {McNichols}, Andrew and {Mehringer}, David and {Miel}, Renaud and {Moellenbrock}, George and {Montesino}, Federico and {Nakazato}, Takeshi and {Ott}, Juergen and {Petry}, Dirk and {Pokorny}, Martin and {Raba}, Ryan and {Rau}, Urvashi and {Schiebel}, Darrell and {Schweighart}, Neal and {Sekhar}, Srikrishna and {Shimada}, Kazuhiko and {Small}, Des and {Steeb}, Jan-Willem and {Sugimoto}, Kanako and {Suoranta}, Ville and {Tsutsumi}, Takahiro and {van Bemmel}, Ilse M. and {Verkouter}, Marjolein and {Wells}, Akeem and {Xiong}, Wei and {Szomoru}, Arpad and {Griffith}, Morgan and {Glendenning}, Brian and {Kern}, Jeff},
        title = "{CASA, the Common Astronomy Software Applications for Radio Astronomy}",
      journal = {\pasp},
         year = 2022,
        month = nov,
       volume = {134},
       number = {1041},
          eid = {114501},
        pages = {114501},
          doi = {10.1088/1538-3873/ac9642},
archivePrefix = {arXiv},
       eprint = {2210.02276},
 primaryClass = {astro-ph.IM},
       adsurl = {https://ui.adsabs.harvard.edu/abs/2022PASP..134k4501C}
}

@ARTICLE{2022AJ....163..135D,
       author = {{de Villiers}, Mattieu S. and {Cotton}, William D.},
        title = "{MeerKAT Primary-beam Measurements in the L Band}",
      journal = {\aj},
         year = 2022,
        month = mar,
       volume = {163},
       number = {3},
          eid = {135},
        pages = {135},
          doi = {10.3847/1538-3881/ac460a},
archivePrefix = {arXiv},
       eprint = {2202.02101},
 primaryClass = {astro-ph.IM},
       adsurl = {https://ui.adsabs.harvard.edu/abs/2022AJ....163..135D}
}

@ARTICLE{2023AJ....165...78D,
       author = {{de Villiers}, Mattieu S.},
        title = "{MeerKAT Holography Measurements in the UHF, L, and S Bands}",
      journal = {\aj},
         year = 2023,
        month = mar,
       volume = {165},
       number = {3},
          eid = {78},
        pages = {78},
          doi = {10.3847/1538-3881/acabc3},
archivePrefix = {arXiv},
       eprint = {2301.06752},
 primaryClass = {astro-ph.IM},
       adsurl = {https://ui.adsabs.harvard.edu/abs/2023AJ....165...78D}
}

@ARTICLE{1997PASP..109..166C,
       author = {{Condon}, J.~J.},
        title = "{Errors in Elliptical Gaussian Fits}",
      journal = {\pasp},
         year = 1997,
        month = feb,
       volume = {109},
        pages = {166-172},
          doi = {10.1086/133871},
       adsurl = {https://ui.adsabs.harvard.edu/abs/1997PASP..109..166C}
}

@ARTICLE{2023ApJS..267....6N,
       author = {{Nita}, Gelu M. and {Fleishman}, Gregory D. and {Kuznetsov}, Alexey A. and {Anfinogentov}, Sergey A. and {Stupishin}, Alexey G. and {Kontar}, Eduard P. and {Schonfeld}, Samuel J. and {Klimchuk}, James A. and {Gary}, Dale E.},
        title = "{Data-constrained Solar Modeling with GX Simulator}",
      journal = {\apjs},
         year = 2023,
        month = jul,
       volume = {267},
       number = {1},
          eid = {6},
        pages = {6},
          doi = {10.3847/1538-4365/acd343},
archivePrefix = {arXiv},
       eprint = {2301.00795},
 primaryClass = {astro-ph.SR},
       adsurl = {https://ui.adsabs.harvard.edu/abs/2023ApJS..267....6N}
}

@ARTICLE{2025FrASS..1266743K,
       author = {{Kansabanik}, Devojyoti and {Gouws}, Marcel and {Patra}, Deepan and {Vourlidas}, Angelos and {Kotz{\'e}}, Pieter and {Oberoi}, Divya and {Shaik}, Shaheda Begum and {Buchner}, Sarah and {Camilo}, Fernando},
        title = "{Solar observation with MeerKAT: demonstration of technical readiness and initial science highlights}",
      journal = {Frontiers in Astronomy and Space Sciences},
         year = 2025,
        month = sep,
       volume = {12},
          eid = {1666743},
        pages = {1666743},
          doi = {10.3389/fspas.2025.1666743},
archivePrefix = {arXiv},
       eprint = {2508.17983},
 primaryClass = {astro-ph.SR},
       adsurl = {https://ui.adsabs.harvard.edu/abs/2025FrASS..1266743K}
}

@ARTICLE{2025MNRAS.540.1685T,
       author = {{Tian}, J. and {Pastor-Marazuela}, I. and {Rajwade}, K.~M. and {Stappers}, B.~W. and {Shaji}, K. and {Hanmer}, K.~Y. and {Caleb}, M. and {Bezuidenhout}, M.~C. and {Jankowski}, F. and {Breton}, R. and {Barr}, E.~D. and {Kramer}, M. and {Groot}, P.~J. and {Bloemen}, S. and {Vreeswijk}, P. and {Pieterse}, D. and {Woudt}, P.~A. and {Fender}, R.~P. and {Wijnands}, R.~A.~D. and {Buckley}, D.~A.~H.},
        title = "{MeerKAT discovery of a hyperactive repeating fast radio burst source}",
      journal = {\mnras},
         year = 2025,
        month = jun,
       volume = {540},
       number = {2},
        pages = {1685-1700},
          doi = {10.1093/mnras/staf793},
archivePrefix = {arXiv},
       eprint = {2505.08372},
 primaryClass = {astro-ph.HE},
       adsurl = {https://ui.adsabs.harvard.edu/abs/2025MNRAS.540.1685T}
}
\bibliographystyle{aasjournalv7}

\end{CJK*}
\end{document}